\documentclass[showpacs,preprintnumbers,amsmath,amssymb,prb,twocolumn,superscriptaddress]{revtex4-1}
\usepackage{amsfonts}
\usepackage{physics}
\usepackage[english]{babel}
\usepackage[T1]{fontenc}
\usepackage{times}
\usepackage{mathrsfs}
\usepackage{graphicx}
\usepackage{dcolumn}
\usepackage{bm}
\usepackage[colorlinks,bookmarks=false,citecolor=blue,linkcolor=red,urlcolor=blue]{hyperref}

\newcommand{\be}{\begin{eqnarray}}
\newcommand{\ee}{\end{eqnarray}}
\newcommand{\ba}{\begin{align}}
\newcommand{\ea}{\end{align}}

\usepackage{tabularx}

\usepackage[usenames,dvipsnames]{xcolor}
\usepackage{soul}
\setulcolor{BrickRed}
\setstcolor{purple}

\makeatletter
\newcommand*\dashline{\rotatebox[origin=c]{90}{$\dabar@\dabar@\dabar@$}}
\makeatother

\begin{document}
	
	\title{Relation between scattering matrix topological invariants and conductance in Floquet Majorana systems}
	
	\author{Thomas Simons}
	\affiliation{Department of Physics, Lancaster University, Lancaster LA1 4YB, United Kingdom}
	\author{Alessandro Romito}
	\affiliation{Department of Physics, Lancaster University, Lancaster LA1 4YB, United Kingdom}
	\author{Dganit Meidan}
	\affiliation{Department of Physics, Ben-Gurion University of the Negev, Beer-Sheva 84105, Israel}
	
	\begin{abstract}
We analyze the conductance of a one-dimensional topological superconductor periodically driven to host Floquet Majorana zero-modes for different configurations of coupling to external leads. We compare the conductance of constantly coupled leads, as in standard transport experiments, with the stroboscopic conductance of pulsed coupling to leads used to identify a scattering matrix topological index for periodically driven systems. 
We find that the sum of DC conductance at voltages multiples of the driving frequency is quantitatively close to the stroboscopic conductance at all voltage biases. This is consistent with previous work  which  indicated that the summed conductance at zero/pi resonance is quantized. We quantify the difference between the two in terms of the width of their respective resonances and analyze that difference for two different stroboscopic driving protocols of the Kitaev chain. While the quantitative differences are protocol-dependent, we find  that generically the discrepancy  is larger when the  zero mode weight at the end of  the chain depends strongly on  the offset time between the driving cycle and the pulsed coupling period.  

	\end{abstract}
	
	\maketitle
	\vspace{.5pc}
	
	\section{Introduction}

Driven non-equilibrium quantum systems can host a variety of distinct phases with no counterparts in equilibrium systems \cite{Khemani2016,Else2016,Yao2017,Zhang2017}. Unlike time independent systems whose properties are intrinsic to the setup and hard to change in situ, the nature of phases in driven systems can be controlled by the more versatile external drive. Of particular interest are topological systems known to host conducting states at the edges of an insulating bulk, which are robust to local disorder and protected by the symmetries of the given system \cite{Hasan2010,Qi2011}.  Subjecting such systems to a source of periodic driving results in the emergence of additional topological phases \cite{Kitagawa2010,Rudner2013,VonKeyserlingk2016,Else2016,Potter2016,Roy2016, Fulga2016,Bauer2019}. One such example is that of a one-dimensional p-wave superconductor (Kitaev chain) subject to a periodic driving of period $T$, which has been shown to possess, in addition to Majorana zero modes at zero energy \cite{Kitaev2001}, protected modes at energy $\pi/T$ (Majorana $\pi$ modes) \cite{Jiang2011,Kundu2013,Farrell2015,Farrell2016,Reynoso2013,Thakurathi2013,Peng2021}. These driven states of matter hold promise  for a wider range of applications \cite{Martin2017,Peng2018,Ozawa2016}. In particular the driven Kitaev wire has been stipulated as a potential candidate for demonstrating a topologically protected, non-Abelian Majorana braiding operation within a single wire \cite{Bauer2019,Bomantara2018,Bomantara2018a}. Such braiding operations are a necessity for topological quantum computing and hence the exploration of diverse alternatives for their realizations is a highly desirable goal.

	The gapless surface states found in topological systems influence the scattering of electrons incident from the leads in an open geometry setup. Scattering matrices provide topological indices for a full classification of topological phases as well as an understanding of the periodicity of the 10 fold way \cite{Fulga2011a,Fulga2012,Meidan2014}. Additionally, expressing the topological indices in terms of a scattering matrix allows one to relate the topology to measurable transport properties. 
Unlike their static  counterparts, the invariants of DC scattering matrices of periodically driven systems are not directly related to the presence of topologically protected   states at their surface. 
	 Instead it is possible to relate  the topological properties  of  Floquet systems to a  {\it gedanken} scattering experiment, in which the leads are coupled to the system only at discrete times separated by the driving period  \cite{Fulga2016}. While the application of these results  to  realistic measurable DC conductance in electronic systems remains unclear  \cite{Moskalets2002b,Perez-Piskunow2014,FoaTorres2014,Kitagawa2011}, 
Floquet topological system have been shown to exhibit a  quantized  {\it sum of conductances} at bias multiples of the driving frequency \cite{Kundu2013,Farrell2015,Farrell2016},  indicating that a modified relation between the two seemingly different physical processes might still exist. 
	

In this manuscript we explore the relationship between topological invariants and the average DC conductance properties of a driven non-interacting electronic system. By analyzing the conductance associated with pulsed coupling to the leads, which defines the scattering matrix topological invariant, we show that it generically differs from its counterpart in the constantly coupled leads setup, i.e. the conductance summed over all Floquet sideband energies, which is accessible in experiments. 
Interestingly, however, we show that, in the limit of small coupling to the leads, the difference between the two is generically small and can be quantified in terms of the difference of the respective resonance widths. We analyze these features for a one-dimensional topological superconductor with two different stroboscopic drivings characterized by sudden switches between two Hamiltonians, $H_0$ and $H_1$ and explore the dependence of the difference between the two conductances on the  different regions of the phase diagram.

\section{Conductance and scattering matrix for different system-lead couplings}
\label{scattering}

\begin{figure}
	\centering
	\includegraphics[width=0.45\textwidth]{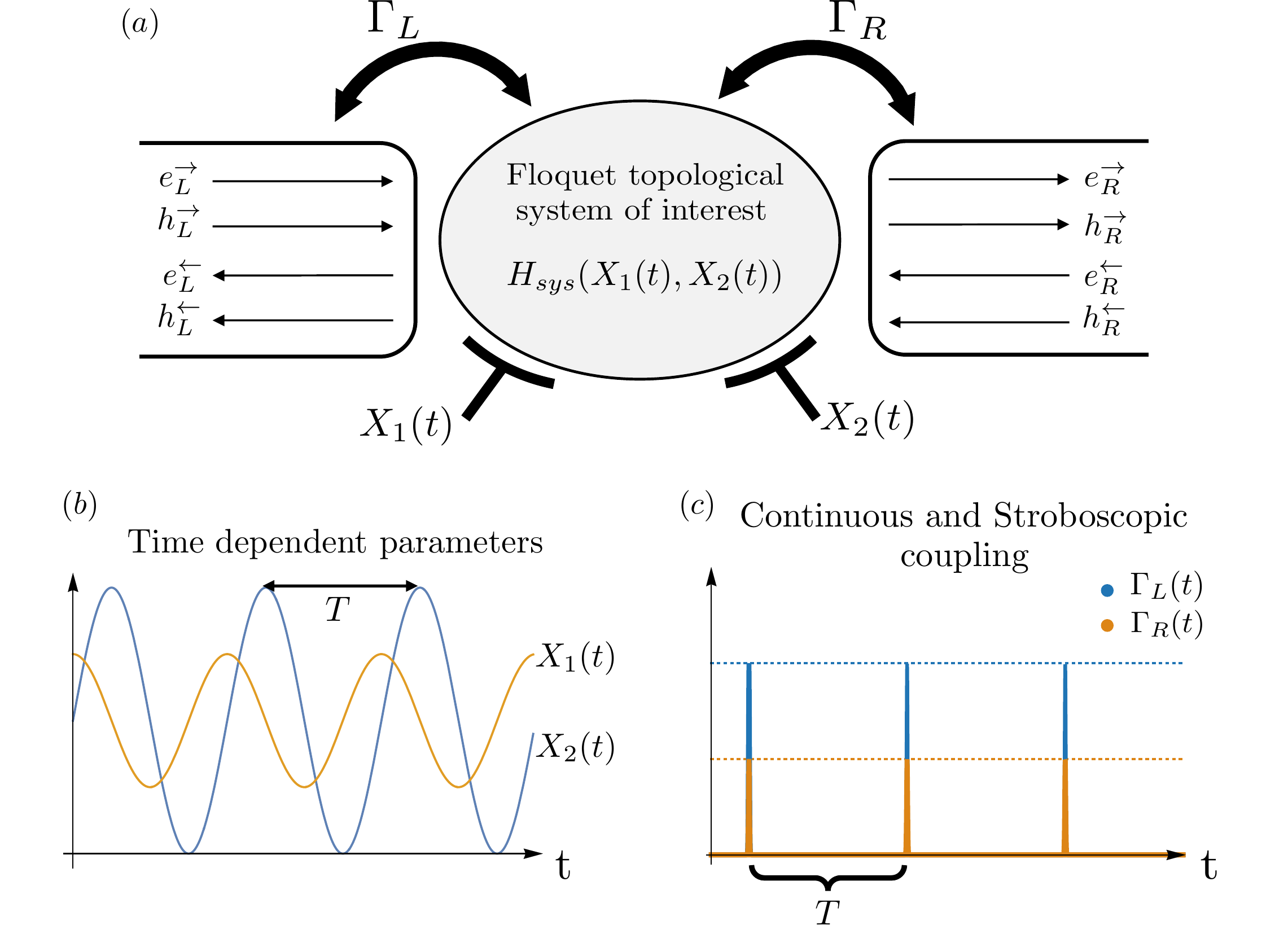}
	\caption{(a) Schematic of an electronic system connected to two external leads (terminals) via tunneling rates $\Gamma_L, \Gamma_R$ and driven via periodic control of its parameters $X_1, X_2$ the time dependence of which is sketched in (b). Each terminal ($L, R$) includes ingoing and outgoing ($\leftarrow, \rightarrow$) electron ($e$) and hole ($h$) scattering states. (c) Two scattering scenarios are depicted, corresponding to either a continuous coupling to the leads (dashed lines) or time-pulsed couplings with periodicity $T$ (solid lines).}
	\label{fig:system}
\end{figure}

We consider a general setup in which a one-dimensional, non-interacting, periodically driven electronic system is in contact with external leads. The latter are electron reservoirs, with constant density of states, tunnel coupled to the system of interest. The system is schematically presented in Fig.\ref{fig:system} and is described by the Hamiltonian
\begin{equation}
H= H_{\rm sys} (t) + H_{\rm T} + H_{\rm lead},
\label{eq:hamiltonian}
\end{equation}
where $H_{\rm{sys}}(t)=\sum_{j,k=1}^N c_j^\dagger h_{j,k} (t) c_k$ is the Hamiltonian of the system of interest, which is assumed to be a generic non interacting system of local fermionic degrees of freedom annihilated by the operator $c_j$. 
The Hamiltonian is explicitly dependent on time in a periodic way, $H_{\rm{sys}}(t+T)=H_{\rm{sys}}(t)$, so that $T=2\pi/\omega$ is the period of the external driving. 
The tunneling to the leads is described by $H_T=\sum_{\alpha,k} \left[\sqrt{\Gamma_\alpha} a_{\alpha,k}^\dagger K_\alpha c_{j_{\alpha}}  + h.c.\right]$, where $K_\alpha$ is the contact matrix between the system and the lead $\alpha$, with $\Gamma_\alpha$ characterizing the coupling strength, and $c_{j_{\alpha}}$ annihilates a particle in the mode $j_\alpha$~\footnote{In a system modeled by spatially discretized sites, $j_\alpha$ labels the spatial coordinate of the system's site closest to lead $\alpha$}. 
Finally, the leads are generic free particle reservoirs with constant density of states and linear dispersion, $H_\alpha= v_\alpha k \sum_k [a_{\alpha,k}^\dagger a_{\alpha,k} - b_{\alpha,k}^\dagger b_{\alpha,k} ]$, where $a_{\alpha,k}$ and $b_{\alpha,k}$ annihilate ingoing and outgoing particles with momentum $k$ in the reservoir $\alpha=L,R$.
The creation/annihilation operators for energy eigenstates in each lead can be identified with the momentum creation operators, e.g.  $b_{L}(E) \equiv b_{L,k}$ and $a_{L}(E)=a_{L,k}$ via $E=\pm v_L k$ respectively. 
 
The transfer of particles between the leads and the system is described by the scattering of ingoing particles  to the outgoing modes of the leads due to the system. The  current in lead $\alpha$ is given by the net flux of particles through a section of the lead at a given position. As long as the energies involved in the transport and driving are much smaller than the Fermi energy of the terminals, the current is expressed in terms of creation and annihilation operators of the scattering states by \cite{doi:10.1142/p822}
\begin{eqnarray}
\label{current def}
I_\alpha(t)= \frac{e}{h}\langle b_\alpha^\dag (t ) b_\alpha (t )\rangle -\langle a_\alpha^\dag (t ) a_\alpha (t )\rangle,
\label{eq:current1}
\end{eqnarray}
where $a_\alpha(t)=(1/2\pi)  \int dt \, e^{iEt} a_\alpha(E)$, $b_\alpha(t)= (1/2\pi) \int dt \, e^{iEt} b_\alpha(E)$, the spatial dependence is immaterial, i.e. $b_\alpha(t) \equiv b_\alpha(x,t)$, and the operators are explicitly time-dependent due to the time-dependence of the system.
Note that, consequently, the current is explicitly time dependent.  It can be expressed in term of the scattering matrix of the system via:
\begin{eqnarray}\label{def:tdepS}
b_\alpha(t) = \int dt' S_{\alpha,\beta}(t,t')a_\beta(t').
\end{eqnarray}
The form of the scattering matrix, and hence the current depends on the system Hamiltonian and on the coupling to the leads.

Rewriting the above  relation in the  energy domain,
we obtain
\begin{eqnarray}
\label{eq:outgoing}
b_\alpha(E) =\sum_{n,\beta} S_{\alpha\beta}(E,E_n) a_\beta(E_n),
\end{eqnarray}
where $E_n \equiv E+ n \omega$ and we have used the periodicity of the system to  constrain the Fourier expression of the scattering matrix to take the form \cite{Moskalets2002b},
\begin{eqnarray}
S(t,t') = \sum_n\int dE e^{-i E(t-t') +in\omega t'}S(E,E_n).
\end{eqnarray}
The two-energy scattering matrix, $S(E,E_n)$, is known as the Floquet scattering matrix, and describes scattering processes in which an outgoing particle of energy $E$ emerges after absorbing/releasing energy in quanta of $\omega$ in the scattering process, and encodes all the information of the time-dependent scattering process \cite{Moskalets2014,Moskalets2004,Blanter2000}.

 Using the Floquet scattering matrix, one can express the current averaged over the driving period $I_\alpha =\frac{1}{T}\int dt I_\alpha(t) $ in the form 
	\begin{equation}
	\label{current}
	\begin{aligned}
	I_\alpha = \frac{e}{h}\int_{-\infty}^{\infty} dE \sum_{\beta \neq \alpha} &\sum_{n}  \Bigg[ \abs{S_{\alpha \beta}(E_n,E)}^2 f_\beta (E)
	\\
	 & - \abs{S_{\beta,\alpha}(E_n,E)}^2 f_\alpha (E)\Bigg],
	\end{aligned}
	\end{equation}
where $f_\alpha(E) = \langle a_\alpha^\dag (E) a_\alpha (E )\rangle$ is the distribution function of particles entering the scatterer through channel $\alpha$. The differential conductance 
is then found via the derivative of the current with respect to the voltage bias, $ G_\alpha = dI_\alpha / dV_\alpha$. 

The formulation can be kept general for non-interacting systems to include superconductors. This requires accounting for both particle and hole degrees of freedom \cite{Beenakker1997}. In this way, the expression for current in Eq. \eqref{current} remains unchanged with the index $\beta$ now running over both particle and hole degrees of freedom in each external lead.
	
For simplicity we consider two external leads with a symmetric voltage bias with respect to the superconductor fermi level, so that the corresponding Fermi distribution functions for both electrons and holes are then given by: $f_{L^e} = f(E-eV) $,  $f_{L^h} = f(E+eV) $, $f_{R^e} = f(E+eV)$ and $f_{R^h}=f(E-eV)$. In the zero temperature limit, the derivative of the Fermi distribution functions are simply step function and hence, after integrating over energy, the contribution to the conductance from each element of the scattering matrix can be expressed as
\begin{equation}
G_{\alpha \beta}(V) = \frac{e^2}{h}\sum_n |{S}_{\alpha \beta}(V_n,V)|^2,
\end{equation}
and the total conductance, in the left lead for example, reads
	\begin{equation}
\label{eq:generic}
	\begin{aligned}
	G_L(V) = & {G}_{L^e L^h}(-V) +  {G}_{L^e R^e}(-V) 
	\\
	&- {G}_{L^e R^h}(V) + \sum_{\beta \neq \alpha}{G}_{\beta L^e}(V). 
	\end{aligned}
	\end{equation}

Eq. (\ref{eq:generic}) is a general expression for a periodically driven electronic systems. In the following, we specialize to two cases in which either (i) we keep constant coupling to the leads as typical in transport measurements setups, or (ii) the system-leads coupling is $\delta$-like pulsed, a configuration of interest in determining the scattering matrix topological invariants{\tiny } of the system.

\subsection{Conductance with constant couplings}

We start by deriving the expression for the conductance for the case in which the coupling to the leads is kept constant and the time dependent driving is applied to the bulk system in order to induce a topological phase. This is the most straightforward setup to be used in transport experiments~\cite{Kundu2013}. 
In order to evaluate  the Floquet scattering matrix for a given scattering system it is useful to first consider the Floquet operator which is  the evolution over a full period $F = U(T,0)$, where the evolution is dictated by an effective (non-Hermitain) Hamiltonian $H_{\rm{sys}}(t) - i\Sigma$, where $\Sigma=\frac{1}{2}\sum_\delta \Gamma_\delta K_\delta{\dagger}K_\delta$ is a self energy accounting for the coupling between the system of interest and the external leads. 

The Floquet operator can be decomposed in terms of left and right eigenstates:
\begin{equation}
\label{Floquet_eigenstates}
\begin{aligned}
	F\ket{\psi_\alpha} = e^{-i(\epsilon_\alpha-i\gamma_\alpha)T}\ket{\psi_\alpha},
	\\
		\bra{\tilde{\psi}_\alpha}F = e^{-i(\epsilon_\alpha-i\gamma_\alpha)T} \bra{\tilde{\psi}_\alpha}.
\end{aligned}
\end{equation}
Here $\epsilon_\alpha$ give the so-called quasienergies of the periodically driven system and are defined modulo the driving frequency $\omega$. The eigenstates of the Floquet operator can be used to define the periodic Floquet eigenstates of the effective Hamiltonian
\begin{equation}
\begin{aligned}
\ket{\Psi_\alpha(t)} = e^{i(\epsilon_\alpha-i\gamma_\alpha)T}U(t,0)\ket{\psi_\alpha},
\\
\bra{\tilde{\Psi}_\alpha(t)} = e^{-i(\epsilon_\alpha-i\gamma_\alpha)T} \bra{\tilde{\psi}_\alpha}U(0,t),
\end{aligned}
\end{equation}
with harmonics given by the Fourier transform
\begin{equation}
\begin{aligned}
\ket{\Psi_\alpha^{(p)}} = \frac{1}{T}\int_{0}^{T} dt e^{ip\omega t } \ket{\Psi_\alpha(t)},
\\
\bra{\tilde{\Psi}_\alpha^{(p)}} = \frac{1}{T}\int_{0}^{T} dt e^{-ip\omega t } \bra{\tilde{\Psi}_\alpha(t)}.
\end{aligned}
\end{equation}
 The harmonics of the Floquet states can be subsequently used to find the Floquet-Green's function $\mathcal{G}^{p}(E)$ \cite{Fruchart2016}:
 \begin{equation}
 \begin{aligned}
 \mathcal{G}^{p}(E) = \sum_{r,\alpha} \frac{\ket{\Psi^{(p+r)}_\alpha}\bra{\tilde{\Psi}^{(r)}_\alpha}}{E-[\epsilon_\alpha+r\omega - i\gamma_\alpha]}.
 \end{aligned}
 \end{equation} 
 From this Floquet-Green's function the scattering matrix elements required to find the conductance across the scattering center can be found via the relation \cite{Arrachea2006}
 \begin{equation}
 S_{\alpha,\beta}(E_m,E_n) = \delta_{\alpha,\beta}\delta_{m-n,0} - i \sqrt{\Gamma_\alpha \Gamma_\beta} \mathcal{G}_{j_\alpha,j_\beta}^{m-n}(E_n),
\label{eq:gf_to_scattering}
 \end{equation}
 where $\Gamma_{\alpha,\beta}$ denote the coupling strength between the system and corresponding scattering channel and $j_{\alpha,\beta}$ labels the mode of the system which is tunnel coupled to lead $\alpha$.
 
In this setup with constant couplings to leads, it has been shown \cite{Kundu2013} that in the presence of a Majorana zero/$\pi$ mode, the  conductance in Eq. \eqref{eq:generic} at resonance  is subject to the quantization condition 
\begin{equation}
\sum_m G_L(\epsilon_{0/\pi}+m \omega)=\frac{2 e^2}{h},
\label{eq:conduttanza_quantizzazione}
\end{equation}
where $\epsilon_0=0 $ and $\epsilon_\pi= \frac{\pi}{T} = \omega/2 $. We will re-derive this result in the following after comparing it with a different setup in which the coupling to the leads is controlled in periodic time-pulses.

\subsection{Stroboscopic Scattering Configuration }

Unlike  the static case, in which the conductance quantization is a direct consequence of the topological index associated with the scattering matrix of the system, there is no direct relation  between the Floquet scattering matrix in Eq. \eqref{eq:gf_to_scattering} and the topological  indices of the driven system. 
Instead, an alternative formulation of the topological index in Floquet systems has been put forward in  terms of a {\it gedanken} scattering configuration,  not immediately  related to measurable quantities \cite{Fulga2016}.  
For a system driven periodically with period $T$, the {\it gedanken} scattering configuration consists of instantaneously emitting and absorbing particles from the leads into the systems at intervals of period $T$. Consequently, the system will evolve with a {\it unitary } Floquet operator $F_t$ between two subsequent couplings to the leads at times $t$ and $t+T$. 
Since the Floquet operator $F_t$ determines the  unitary evolution of the decoupled  system over a period, the corresponding  stroboscopic  scattering matrix is expressed  as ~\cite{Fulga2016, Fyodorov2000}
\begin{equation}
\label{FulgaSmtx}
\begin{aligned}
S_t^{\rm{strob}}(E)=&\sqrt{\mathcal{I}-  W W^\dagger} - 
\\
&W \frac{1}{\mathcal{I} - e^{iET}F_t\sqrt{\mathcal{I}- W^\dagger W}}e^{iET}F_tW^\dagger,
\end{aligned}
\end{equation}
where the matrix $W$ encodes the coupling to the leads and, using the notation from Eq. \eqref{eq:hamiltonian}, takes the form:
$
W=\sum_\delta \sqrt{T \Gamma_\delta} K_\delta.
$

Physically, the scattering matrix in Eq. \eqref{FulgaSmtx} describes the  situation in which particle scatter into the absorbing terminals stroboscopically, only at the beginning and end of each time period \cite{Fyodorov2000,Ossipov2002}. Specifically, once  an arbitrary starting point $t=0$ has been set for the periodic driving, we assume that couplings to the terminals  are performed  by omitting and collecting  particles at times that are separated by full periods at a time offset of $t$ with respect to the driving period. If we define the Floquet operator $F \equiv U(T,0)$, the stroboscopic  scattering matrix is obtained using
	\begin{equation}
	F_t = U(t,0) F U^\dagger(t,0),
	\end{equation}
corresponding to the Floquet evolution operator which explicitly depends on the offset time $ t$. Note that, while the spectrum of the Floquet operator $F_t$, which discriminates the existence of Floquet edge states, is of course independent of the offset time $t$, the stroboscopic scattering matrix depends in principle on $t$. In fact, physically, the scattering between the times $t$ and $t+T$ depends on the specifics of the evolution between the two times, and hence on $t$ itself.

This stroboscopic scattering matrix can be used to define a corresponding conductance that is averaged over the offset time $t$, $I^{{\rm{strob}}}_\alpha = \frac{1}{T} \int_0^T dt I^{\mathrm{strob}}_{t,\alpha}$, where
\begin{equation}
\label{Fulgacurrent}
\begin{aligned}
I^{\mathrm{strob}}_{t,\alpha} = \frac{e}{h}\int_{-\infty}^{\infty} dE& \sum_\beta \Bigg[ \abs{S_{t,\alpha \beta}^{\rm{strob}}(E)}^2 f_\beta (E) 
\\
&- \abs{S_{t,\beta \alpha}^{\rm{strob}}(E)}^2 f_\alpha (E)\Bigg].
\end{aligned}
\end{equation}
The corresponding contribution to the stroboscopic conductance from each element of the scattering matrix at zero temperature is then given by
\begin{equation}
\label{strob cond comp}
G^{\rm{strob}}_{\alpha \beta}(V) = \frac{e^2}{h} \frac{1}{T}\int_0^T dt \abs{S_{t,\alpha \beta}^{\rm{strob}}(V)}^2,
\end{equation}
and the total conductance reads
\begin{equation}
\label{eq:strobcond}
\begin{aligned}
G_L^{\rm{strob}}(V) =  {G}^{\rm{strob}}_{L^e L^h}(-V) &+  {G}^{\rm{strob}}_{L^e R^e}(-V) 
\\
&- {G}^{\rm{strob}}_{L^e R^h}(V) + \sum_{\beta \neq \alpha}{G}^{\rm{strob}}_{\beta L^e}(V) . 
\end{aligned}
\end{equation}
The voltage profile of this  stroboscopic conductance will reflect the existence of topological indices of the  periodically driven system, which  can be formulated in terms of the corresponding stroboscopic   scattering matrix [Eq. \eqref{FulgaSmtx}]. Namely, the presence of topologically protected edge states will result in quantized conductance peaks. While there is a similarity between the conductance sum-rule quantization in Eq. \eqref{eq:conduttanza_quantizzazione} and the quantization of the pulsed conductance in Eq. \eqref{eq:strobcond}, the relationship between these two quantities remains to be explored. 

Unlike the physical conductance for a system continuously coupled to the external leads [Eq. (\ref{eq:generic})], Eq. \eqref{FulgaSmtx} shows that the stroboscopic scattering matrix and consequently the  stroboscopic current are both periodic in energy. This is consistent with the fact that the quantization condition for the physical conductance is expressed as a sum rule Eq. \eqref{eq:conduttanza_quantizzazione}. This relation is further highlighted by considering  an intermediate scenario  in which the system is continuously coupled to the external leads but outgoing scattered particles are only measured at discrete times separated by the driving period and starting at the delay time $t$. 

In this intermediate scenario, the relation between the stroboscopic scattering matrix with continuous coupling to the leads, denoted $\tilde{S}_t(E)$, and $S(E,E_n)$ can now be obtained by explicitly considering that the scattering operators are relevant only at discrete times, hence they can be expanded in a Fourier series
\begin{eqnarray}
a(lT+t) = \int dE \tilde{a}(E) e^{-i E(lT+t)},
\end{eqnarray}
where $\tilde{a}(E) = \sum_l  a(lT+t) e^{i E(lT+t)}$ and $\langle {\tilde{a}_\alpha}^\dag (E') \tilde{a}_\alpha (E )\rangle = \tilde{f}_\alpha (E) \delta(E-E')$. Using the Fourier expansion of the scattering matrix in Eq. \eqref{eq:current1}, (the details of the computation are presented in Appendix \ref{app:appendice}), we find that
\begin{equation}
\label{eq:corrente}
I_\alpha(lT+t)=\frac{e}{h}\!\!\int \!\!dE\sum_{\beta} \left[ |\tilde{S}_{t,\alpha\beta}(E)|^2\tilde{f}_\beta(E)\right]-\tilde{f}_\alpha (E), 
\end{equation}
and $\tilde{S}_t(E)$, can be expressed in terms of the Floquet scattering matrix via
	\begin{equation}
	\tilde{S}_t(E) = \sum_{n,k} S(E_n,E_{n+k})e^{ik \Omega t}.
	\end{equation}  
	
Although  $\tilde{S}_t(E)$ fails to include scattered particles in between the stroboscopic detection times and is hence a non-unitary construction, this relationship to the Floquet scattering matrix highlights the need to sum scattering events over all Floquet sideband energies, $E_n = E+n\omega$, in order to make a meaningful comparison. 
We therefore define a summed  conductance
\begin{equation}\label{sum_conductance}
\tilde{G}_\alpha(V) = \sum_n G_\alpha(V+n \omega),
\end{equation}
which reproduces the expression in Eq. \eqref{eq:conduttanza_quantizzazione} and provided us with the natural quantity to be compared with the quantized conductance at zero bias in Eq. \eqref{eq:strobcond}.

\subsection{Weak Coupling Limit}

Since both $\tilde{G}_\alpha(V)$ and $G^{\rm{strob}}_\alpha(V)$ are expected to display a peak at $V=\epsilon_{0/\pi}$ with the same quantized maximum, we can analyze their behavior  in the vicinity of the resonance  $|eV-\epsilon_{0/\pi}| \ll \omega$ and at  small coupling (which is expected to set the width of the peak) to appreciate the difference between the two.
In the case in which the system-lead coupling can be considered weak with respect to the other energy scales associated with the system Hamiltonian $H_{\rm sys}(t)$, the self energy contribution $\Sigma$ can be treated as a perturbation. In particular, the Floquet states appearing in the definition of the Floquet-Green's function, can be approximated by the solutions to the uncoupled  Floquet equation,
\begin{equation}
\Big(H_{\rm sys}(t) - i \frac{d}{dt} \Big) \ket{\phi_\alpha(t)} = \epsilon_\alpha \ket{\phi_\alpha(t)}.
\end{equation} 
The first order corrections to the quasienergies due to the perturbation are found to be
\begin{equation}
\tilde{\gamma}_\alpha = \frac{1}{T} \int_{0}^{T} dt \expval{\Sigma}{\phi_\alpha(t)},
\end{equation}
with the self energy term again defined as $\Sigma=\frac{1}{2}\sum_\delta \Gamma_\delta K_{\delta}^{\dagger}K_\delta$. The elements of the Floquet scattering matrix that contribute to the current in Eq. \eqref{current} can then be approximated as
\begin{widetext}
\begin{equation}
\begin{aligned}
\abs{S_{\alpha \beta}(E_n,E)}^2 =
\Gamma_\alpha \Gamma_\beta \sum_{i j r r'} \frac{\bra{\alpha} K_\alpha \ket{\phi_i^{(n+r)}} \bra{\phi_i^{(r)}} K^\dagger_\beta \ket{\beta}\bra{\beta} K_\beta \ket{\phi_j^{(n+r')}}\bra{\phi_j^{(r')}} K^\dagger_\alpha \ket{\alpha}}{\big[E-(\epsilon_i + r \omega - i\tilde{\gamma}_i)\big] \big[E-(\epsilon_j + r' \omega + i\tilde{\gamma}_j)\big]},
\end{aligned}
\end{equation}
\end{widetext}
where  $\bra{\phi_j} K_\alpha \ket{\alpha}$  is the tunnel matrix element between the mode $\phi_j$ of the system and the scattering mode $\alpha$, and  $\alpha$ runs over both particle and hole scattering modes in each external lead. 

In the limit of weak coupling, the  elements of scattering matrix take the form of sharp Lorentzian peaks at energies $\epsilon_i + r\omega$ and $\epsilon_j + r'\omega$. Consequently, the sums are dominated by contributions for which these energies coincide and hence $i=j$ and $r=r'$ or for which the quasienergies $\epsilon_i$ are degenerate. Hence, in this limit, the scattering matrix components read 
\begin{equation}
\label{Floquet scat approx}
\begin{aligned}
&\abs{S_{\alpha \beta}(E_n,E)}^2 \approx
\\
& \Gamma_\alpha \Gamma_\beta \sum_{i r} \frac{\abs{\sum_k \bra{\alpha}K_\alpha\ket{\phi_{i_k}^{(n+r)}} \bra{\phi_{i_k}^{(r)}}K_\beta\ket{\beta}}^2}{{\tilde{\gamma}_i}^2+(E-\epsilon_i+r\omega)^2},
\end{aligned}
\end{equation}
where $\ket{\phi_{i_k}}$ represent the eigenstates corresponding to the degenerate eigenvalue $\epsilon_i$. As Majorana  bound states are localized at one end of the chain, they    only couple to one of the external leads. Consequently, their  contribution to the conductance will arise from  from the Andreev reflection components
 of the scattering matrix. This localization of the degenerate Floquet eigenstates allows us to henceforward drop the sum over degenerate states $k$ in expressions for the scattering matrices. 
 
The contribution to the summed conductance $\tilde{G}(V)$ at zero temperature from each term of the scattering matrix reads
\begin{equation}
\label{summed cond comp}
\begin{aligned}
\tilde{G}_{\alpha,\beta}(V) = \frac{e^2}{h}\Gamma_\alpha \Gamma_\beta \sum_{i r n m } \frac{\abs{ \bra{\alpha}K_\alpha\ket{\phi_{i}^{(n)}} \bra{\phi_{i}^{(r)}}K_\beta\ket{\beta}}^2}{{\tilde{\gamma}_i}^2+(V-\epsilon_i+m\omega)^2},
\\
 = \frac{e^2}{h}\frac{\Gamma_\alpha \Gamma_\beta}{T^2} \sum_{i m } \int_0^T dt dt' \frac{\abs{ \bra{\alpha}{K_\alpha \ket{\phi_{i}(t)}} \bra{\phi_{i}(t')}K_\beta \ket{\beta}}^2}{{\tilde{\gamma}_i}^2+(V-\epsilon_i+m\omega)^2}.
\end{aligned}
\end{equation}
Close to the resonant quasi-energies, $V \approx \epsilon_i/e$, the conductance contributions take the form of a Lorentzian distribution:
\begin{equation}
\tilde{G}_{\alpha,\beta}(V) \approx \frac{e^2}{h}\frac{\tilde{\gamma}^{(\alpha)}_{i}\tilde{\gamma}^{(\beta)}_{i}}{\tilde{\gamma}_i^2} \mathcal{L}\Big(\frac{eV-\epsilon_i}{\tilde{\gamma}_i}\Big),
\end{equation}
where $\mathcal{L}(x) = (1+x^2)^{-1}$ is the Lorentzian function and
\begin{equation}
\begin{aligned}
&\tilde{\gamma}_i^{(\alpha)} = \frac{1}{T}\int_0^T \Gamma_\alpha \abs{\bra{\alpha}K_\alpha\ket{\phi_i(t)}}^2, 
\\
&\mathrm{so \ that} \ \ \tilde{\gamma}_i = \sum_\delta \tilde{\gamma}_i^{(\delta)}.
\end{aligned}
\end{equation}
Since the Majorana bound states are localized at one end of the system, they contribute  to the conductance through Andreev reflection only. The particle-hole symmetry of the system also dictates that for Majorana states we have that $\tilde{\gamma}_{0,\pi}^{(L^e)} = \tilde{\gamma}_{0,\pi}^{(L^h)} $. Consequently, for $V\approx\epsilon_{0/\pi}/e$,
\begin{equation}
\label{summed cond at resonance}
\tilde{G}(V) =2\tilde{G}_{L^e,L^h}(V) \approx \frac{2e^2}{h}\mathcal{L}\Big(\frac{eV-\epsilon_{0/\pi}}{\tilde{\gamma}_{0/\pi}}\Big).
\end{equation}

Next we derive the expression for the stroboscopic conductance in the vicinity of the resonances.  In order to compare these two quantities, it is instructive to express the stroboscopic conductance in terms of the Floquet eigenstates of the unperturbed driving Hamiltonian $H_{\rm sys}(t)$. We would like to use perturbation theory to perform an expansion in the coupling strength, $\Gamma$, of the operator appearing as a fraction in Eq. \eqref{FulgaSmtx}. We can first rewrite this expression as a geometric series,
	\begin{equation}
	\begin{aligned}
	\frac{1}{\mathcal{I} - e^{iET}F_t\sqrt{\mathcal{I}-W^\dagger W}} &= \sum_k \big(e^{iET}F_t\sqrt{\mathcal{I}- W^\dagger W} \big)^k
	\\
	&\approx \sum_k \big(e^{iET}(\underbrace{F_t}_{A_0}- \underbrace{\frac{1}{2}F_tW^\dagger W}_{\Gamma A_1}) \big)^k.
	\end{aligned}
	\end{equation}
	We can write the operator $A = A_0 + \Gamma A_1$ in terms of its eigenstates defined as, $A \ket{x_i} = x_i\ket{x_i}$:
	\begin{equation}
	\begin{aligned}
	\frac{1}{\mathcal{I} - e^{iET}F_t\sqrt{\mathcal{I}-W^\dagger W}} &= \sum_k  \big(e^{iET}\sum_i x_i \ket{x_i}\bra{x_i} \big)^k
	\\
	&=\sum_{i} \frac{1}{1-e^{iET}x_i}\ket{x_i}\bra{x_i}.
	\end{aligned}
	\end{equation}
	The Floquet operator of the decoupled system calculated at a particular offset time $t$ can be expanded in terms of its eigenstates as
	\begin{equation}
	\begin{aligned}
	F_t =  \sum_{i} e^{-i\epsilon_i T}\ket{\phi_i(t)} \bra{\phi_i(t)}.
	\end{aligned}
	\end{equation} 
	The eigenstates and eigenvalues of $A_0$ are hence given by $\ket{\phi_i(t)}$ and $e^{i\epsilon_i T}$ respectively. We can then use perturbation theory to calculate the first order correction to the eigenvalues due to the perturbed operator $A$:
	\begin{equation}
	\label{perturbed energy}
	\begin{aligned}
	x_i(t) &= e^{-i\epsilon_i T} + \frac{1}{2} \frac{\bra{\phi_i(t)}F_tW^{\dagger}W \ket{\phi_i(t)}}{\bra{\phi_i(t)}\ket{\phi_i(t)}}
	\\
	&= e^{-i\epsilon_i T}\Big(1+T\underbrace{\bra{\phi_i(t)} \Sigma \ket{\phi_i(t)}}_{\gamma_i(t)}\Big)
	\end{aligned}
	\end{equation}
    The stroboscopic scattering matrix now takes the form
	\begin{equation}
	\begin{aligned}
	S_t^{\rm{strob}}(E) =& \sqrt{\mathcal{I} - W^\dagger W}
	\\
	-& W \sum_{i} \frac{e^{i(E-\epsilon_i)T}}{1-e^{iET}x_i(t)} \ket{\phi_i(t)}\bra{\phi_i(t)} W^\dagger.
	\end{aligned}
	\end{equation}
	
	Contributions to the stroboscopic conductance defined in Eq. \eqref{strob cond comp} then read
	\begin{equation}
	\begin{aligned}
	G^{\rm{strob}}_{\alpha \beta}(V) = \ \ \ \ \ \ \ \ \  &
	\\
	\frac{e^2}{h}T \Gamma_\alpha \Gamma_\beta\int_0^T dt \sum_{i} &\frac{ \abs{ \bra{\alpha}K_\alpha\ket{\phi_{i}(t)}\bra{\phi_{i}(t)}K_\beta^\dagger \ket{\beta} }^2}{\abs{1-e^{i(V-\epsilon_i)T}(1+T\gamma_i(t))}^2}. 
	\end{aligned}
	\end{equation}
	Here we have again used the fact that, in the limit of weak coupling, the conductance profile will consist of sharp peaks at quasienergies $\epsilon_i$.
In order to directly compare the stroboscopic and Floquet conductances in the weak coupling limit, it is instructive the rearrange the expression for the conductance summed over Floquet sidebands (Eq. \eqref{summed cond comp}) into a similar form:
	\begin{equation}
\label{eq:comparison}
	\begin{aligned}
	\tilde{G}_{\alpha,\beta}(V) = &
	\\
	= \frac{e^2}{h}\Gamma_\alpha \Gamma_\beta &\sum_{i} \int_0^T dt dt' \frac{\abs{ \bra{\alpha}{K_\alpha \ket{\phi_{i}(t)}} \bra{\phi_{i}(t')}K_\beta \ket{\beta}}^2}{\abs{1-e^{i(V-\epsilon_i)T}(1+T\tilde{\gamma_i})}^2},
	\end{aligned}
	\end{equation} 
 where we have used the relation 
\begin{equation}
\sum_p \frac{e^{ip\omega z}}{A-p \omega} = \frac{Tie^{iAz}}{e^{iAT}-1}.
\end{equation}
Eqs. \eqref{eq:comparison} and \eqref{summed cond comp} show similarities between the conductance  obtained for   constant and  stroboscopic coupling to the leads  in the sense that they are both dominated by resonances in the weak coupling limit. Even in the weak coupling limit, the two expressions might remain generically different since the width of the respective resonances is controlled by different parameters. 
	 
In order to identify this difference, we can compare the expressions close to a resonance in the weak coupling limit. Approaching the resonant quasienergies $\epsilon_i$ ($V \approx \epsilon_i/e$), the contributions to the stroboscopic conductance from each scattering matrix element can be further simplified as 
	\begin{equation}
	\begin{aligned}
     &G_{\alpha \beta}^{\mathrm{strob}}(V) =
	\\
	&\frac{e^2}{h}\frac{1}{T}\int_{0}^{T}\frac{\abs{ \bra{\alpha} K_\alpha\ket{\phi_{i}(t)}\bra{\phi_{i}(t)}K_\beta^\dagger \ket{\beta} }^2}{\gamma^2_i(t)} \mathcal{L}\Big(\frac{eV-\epsilon_i}{\gamma_i(t)}\Big)
	\\
	&=  \frac{e^2}{h}\frac{1}{T}\int_{0}^{T}\frac{\gamma^{(\alpha)}_{i}(t)\gamma^{(\beta)}_{i}(t)}{\gamma_i^2(t)} \mathcal{L}\Big(\frac{eV-\epsilon_i}{\gamma_i(t)}\Big),
	\end{aligned}
	\end{equation}
	where
	\begin{equation}\label{t_dep_width}
	\begin{aligned}
	 &\gamma_i^{(\delta)}(t) = \Gamma_\delta \abs{\bra{\delta}K_\delta\ket{\phi_i(t)}}^2, 
	 \\
	 \mathrm{so \ that} \ \ &\gamma_i(t) = \sum_\delta \gamma_i^{(\delta)}(t).
	\end{aligned}
	\end{equation}
Again the localized nature of the Majorana states along with the particle-hole symmetry of the system mean that the conductance at these energies (i.e. $V\approx \epsilon_{0/\pi}/e$) can be expressed as
\begin{equation}
\label{strob cond at resonance}
G^{\rm{strob}}(V) = \frac{2e^2}{h}\frac{1}{T} \int_{0}^{T}dt \mathcal{L}\Big(\frac{eV-\epsilon_{0/\pi}}{\gamma_{0/\pi}(t)}\Big).
\end{equation}

Eqs. \eqref{summed cond at resonance} and \eqref{strob cond at resonance} are the main finding of this section. Their comparison shows that the discrepancy between the two can be quantified by the time dependence of the  function  $\gamma_i(t)$ defined in Eq. \eqref{perturbed energy},  and that the two quantities agree in the case that this function is time-independent.
As a figure of merit that quantifies the discrepancy between the two conductance setups at each resonance, we can introduce the quantity:
	\begin{equation}\label{diff_fun}
	D_{i} = \frac{1}{\Gamma}\int_{\epsilon_{i}-\Gamma}^{\epsilon_{i}+\Gamma}dE\Bigg[\mathcal{L}\Big(\frac{E-\epsilon_{i}}{\tilde{\gamma}_{i}}\Big) -\frac{1}{T} \int_{0}^{T}dt \mathcal{L}\Big(\frac{E-\epsilon_{i}}{\gamma_{i}(t)}\Big)\Bigg].
	\end{equation}
To explore these features we analyze below  the constant coupling conductance and the  pulsed coupling conductance of the periodically driven Kitaev chain subject to two different driving protocols.

\section{Transport signatures of a periodically driven Kitaev chain}
\begin{figure*}
	\centering
	\includegraphics[width=1\textwidth]{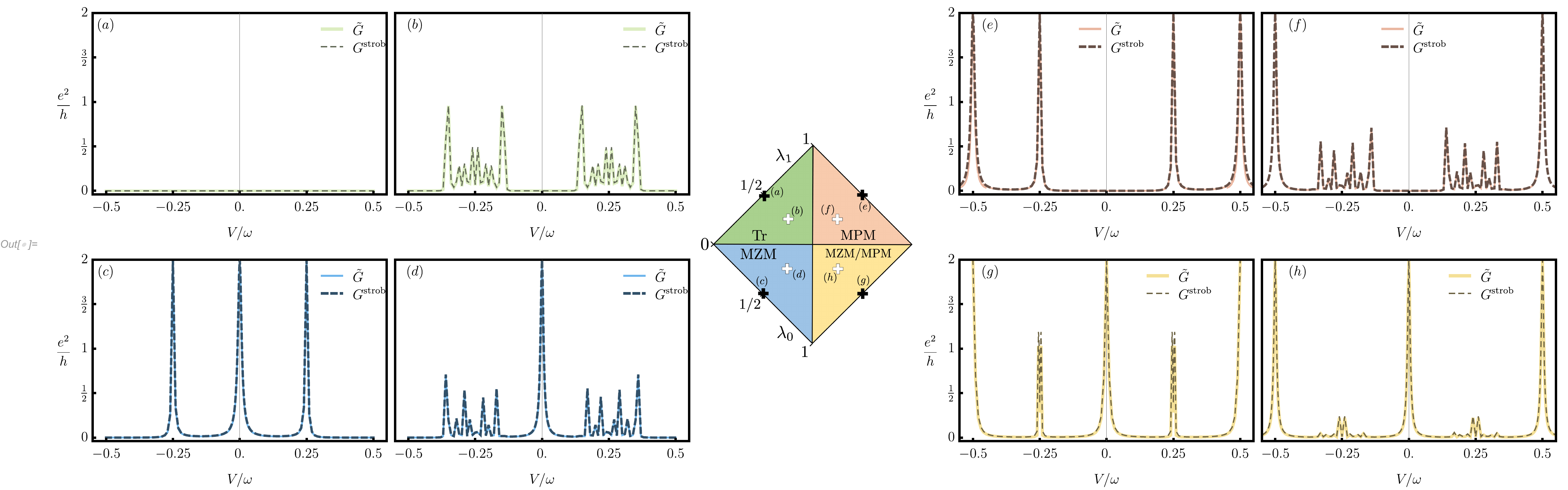}
	\caption{Phase space diagram illustrating how the topological phase of the Floquet Kitaev wire depends upon the Hamiltonian parameters $\lambda_0$ and $\lambda_1$ (see Eqs. \eqref{Kitaev Ham 1} \eqref{Kitaev Ham 2}). (a-h) Numerical results for the zero temperature differential conductance summed of energy sidebands, $\tilde{G}$ and the stroboscopic conductance $G^{\mathrm{strob}}$, plotted as a function of the total voltage bias between the left and right external leads $V = V_L-V_R$. The plot colours correspond to those in the phase diagram. The conductance is evaluated at both the four sweet spots, marked by black crosses and plotted with the darker shade, as well as the points marked by the white crosses in for each phase: (a,b) Trivial, (c,d) MZM, (e,f) MPM, (g,h) MZM/MPM. The results were obtained using a chain of 20 sites and with tunneling rates to the external leads given by $\Gamma_{L/R}/\omega=0.016$.
	}
	\label{fig:zerocond}
\end{figure*}

We demonstrate the difference between the two expressions for the conductance using the driven Kitaev chain, which is  a simple model that displays multiple  topological phases of driven systems. The Hamiltonian for this systems is given by
\begin{eqnarray}
\nonumber
H(w, \Delta,\mu) &=& \sum_i  \left[-\frac{w}{2} c_i^\dag c_{i_+1} +\frac{\Delta}{2} c_i^\dag c_{i_+1}^\dag+ {\rm h.c.} \right]\\
&&-\mu \sum_i  c_i^\dag c_i 
\end{eqnarray}

We consider two different step like driving protocols
in which the parameters of the Hamiltonian are switched instantaneously between two different sets of values. 

\subsection{Sudden switching between Hamiltonians in different topological phases}
In the first protocol,  following Ref. \onlinecite{Bauer2019} we consider the two part driving protocol which switches between a topologically trivial and a topologically non trivial Hamiltonians, expressed by the  Floquet  operator:
\begin{eqnarray}\label{MaresaDriving}
F= e^{-iH_1T/2}e^{-iH_0T/2},
\end{eqnarray}
where 
\begin{eqnarray}
\label{Kitaev Ham 1}
H_0&=&H(2\pi   \lambda_0/ T,2\pi   \lambda_0/ T,0),\\
\label{Kitaev Ham 2}
H_1 &=&H(0,0,2\pi   \lambda_1/ T),
\end{eqnarray}
correspond to the static Hamiltonian of the topological phase and the trivial phase of the Kitaev chain, respectively.  $H_0$ describes  the sweet spot of  the topological phase which is  characterized by Majorana zero  modes, with zero correlation length.  

The phase diagram of driven system is plotted in Fig. \ref{fig:zerocond} in the parameter regime  $0<\lambda_i< 1$. The system exhibits 4 distinct topological phases distinguished by the presence or absence of Majorana zero modes and Majorana $\pi$ modes. The phases  can be identified via the topological index expressed in terms of the scattering matrix \cite{Fulga2016} via
\begin{eqnarray}
\nu_{0/\pi}=\frac{1}{i\pi} \log {\det{{\cal R}_L(\epsilon_{0/\pi})}},
\end{eqnarray}
where $\cal{R}_L$ is the part of the entire stroboscopic scattering matrix that describes reflection in the lead $L$.
 A clear insight can be obtained by analyzing the stroboscopic  scattering matrix  at the sweet spots which are characterized by Majorana modes localized at the left and right most sites, and for perfectly transparent leads $
 W=\sum_\delta \sqrt{T \Gamma_\delta} K_\delta.
 =\sum_\delta K_\delta$. In this limit  there is no transmission and  the scattering matrix decouples into $S(\epsilon)= {\cal R}_L(\epsilon)\oplus {\cal R}_R(\epsilon)$, which are both two by two reflection matrices at the two leads. 

From Eq. \eqref{FulgaSmtx}, we have:
\begin{itemize}
\item[(i)]For the trivial phase sweet spot at  $ \lambda_1=1/2 $ and $\lambda_0=0 $ we have ${\cal R}_L(\epsilon)={\cal R}_R(\epsilon)=- i e^{i\epsilon T} \sigma_z$, and the topological index is:
\begin{eqnarray}
\nu_{0/\pi}=0.  
\end{eqnarray}
\item[(ii)] The MZM phase, where the sweet spot is $\lambda_0 = 1/2  $ and  $\lambda_1  =0 $ results in ${\cal R}_L(\epsilon)=e^{i\epsilon T}/2  [ (1- e^{i\epsilon T}) \boldsymbol{1} - (1+e^{i\epsilon T}) \sigma_x ]$ and ${\cal R}_R(\epsilon)=e^{i\epsilon T}/2  [ (1- e^{i\epsilon T}) \boldsymbol{1} + (1+e^{i\epsilon T}) \sigma_x ]$,  and the topological index is:
\begin{eqnarray}
	\nu_{0}= 1 \,; \,\nu_{\pi}=0.  
\end{eqnarray}
\item[(iii)] The sweet spot of the MPM phase is at $\lambda_0 = 1/2 $ and $\lambda_1  =1 $ with  ${\cal R}_L(\epsilon)=e^{i\epsilon T}/2  [ (-1- e^{i\epsilon T}) \boldsymbol{1}+  (1-e^{i\epsilon T}) \sigma_x ]$ and ${\cal R}_R(\epsilon)=e^{i\epsilon T}/2  [ (-1- e^{i\epsilon T}) \boldsymbol{1} - (1-e^{i\epsilon T}) \sigma_x ]$ ,  and the topological index is:
\begin{eqnarray}
	\nu_{0}= 0 \,; \,\nu_{\pi}=1. 
\end{eqnarray}
\item[(iv)] Finally, the sweet spot of the MPM+MZM phase at $\lambda_0 = 1 $ and $\lambda_1  =1/2 $ gives ${\cal R}_L(\epsilon)=-{\cal R}_R(\epsilon)=e^{i\epsilon T} \sigma_y$.
\begin{eqnarray}
	\nu_{0}= 1 \,; \,\nu_{\pi}=1. 
\end{eqnarray}
\end{itemize}
These results are directly reflected in the quantized values of $G^{\rm strob}(V=\epsilon_{0/\pi})$.


%

Fig. \ref{fig:zerocond}  shows the zero temperature conductance profiles as a function of bias voltage obtained for the driving protocol in Eq. \eqref{MaresaDriving} for each of the 4 different topological phases at the sweet spots (black crosess) and  away from the sweet spots (white crosses). Results were obtained numerically using a chain of 20 fermionic sites. This chain is sufficiently long that, for the parameters studied, any interaction between Majorana modes at either end of the superconductor, and hence any splitting of the Majorana conductance peaks, is negligible. 
The solid line corresponds to the physical conductance of the system, summed over Floquet harmonics, calculated from Eqns. \eqref{eq:generic}  and \eqref{sum_conductance} and the dashed line corresponds to the stroboscopic conductance calculated from the fictitious pulsed scattering problem in Eq. \eqref{Fulgacurrent}. 
The three topological non-trivial phases are characterized by the existence of conductance peaks of height $2e^2/h $ at  $eV = 0$ and/or $eV=\pi/T$, corresponding to the existence of Majorana zero  and $\pi$ modes, respectively.
 These peaks arise due to resonant Andreev reflection events via the Majorana modes and the values of voltage bias at which they occur can be deduced via the maxima of the off-diagonal elements of the corresponding reflection matrices.
  
\begin{figure*}
	\centering
	\includegraphics[width=0.9\textwidth]{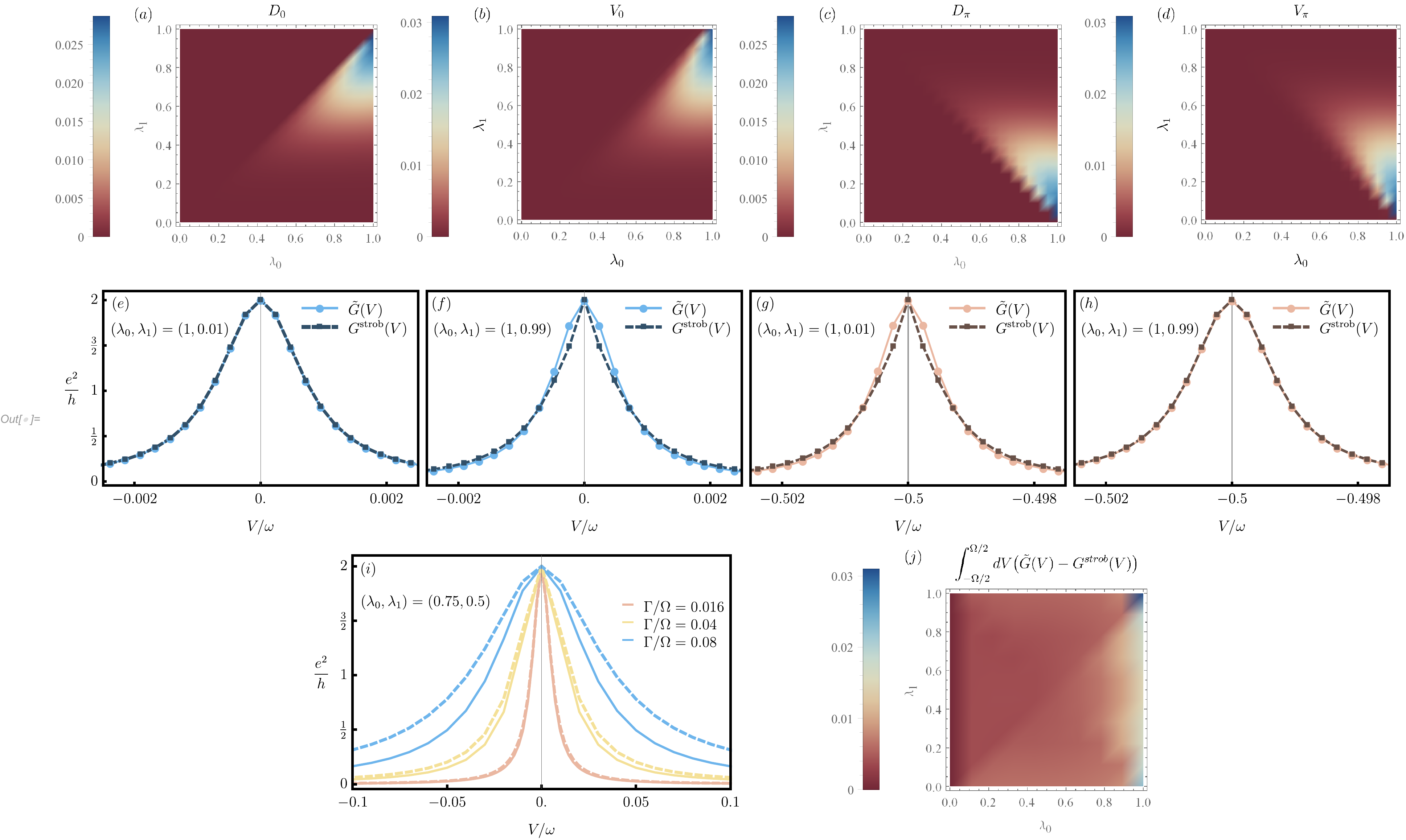}
	\caption{(a,c) Density plots illustrating the value of the difference function for both the zero mode resonance $D_0$ and $\pi$ mode resonance $D_\pi$ respectively throughout the parameter space $(\lambda_0,\lambda_1)$ with coupling strength $\Gamma_{L/R}/\omega=0.0016$. (b,d) Corresponding plots of the time variance of the function $\gamma_\alpha(t)$ controlling the resonance widths. The comparison between the conductance summed over energy sidebands $\tilde{G}$ and the stroboscopic conductance $G^{\mathrm{strob}}$ for selected points are shown in (e-h) again with coupling strength $\Gamma_{L/R}/\omega=0.0016$. (i) Comparison of $\tilde{G}$ (solid lines) and $G^{\rm{strob}}$ (dashed lines) for increased strength coupling to the external leads $\Gamma$ close the MZM resonance. (j) Density plot illustrating difference between the stroboscopic and summed conductances integrated over the entire spectrum throughout the parameter space, calculated using a coupling strength of $\Gamma_{L/R}/\omega=0.016$. All data was obtained using a chain of $n=20$ fermionic sites.}
	\label{fig:OurHDifferences}
\end{figure*}  
As discussed in Sec. \ref{scattering}, the DC conductance and averaged stroboscopic conductance correspond to  different scattering setups with constant vs. pulsed coupling to the leads, and consequently  different broadening of levels. As a result, the respective conductance traces are expected to differ in the width of the resonant conductance peaks. For weak coupling  to the leads we find that  the conductance traces show a very good agreement, even away from the $eV = 0$ and $eV=\pi/T$ peaks ~\footnote{In Fig. \ref{fig:zerocond}, panel (g), the the height of the solid and solid and dashed curves in the bulk differ due to the mixing of the resonances in the bulk. The agreement in the heights is recovered numerically for sufficiently small $\Gamma_{L/R}/\omega= 0.001$, for which the width of the peak could not be displayed on the figure scale.}.
 
In order to quantify the discrepancy  between the two conductances, we plot the difference function $	D_{0/\pi} $ (Eq. \eqref{diff_fun}) around the  $eV = 0$ and $eV = \pi/T$ resonances  throughout the parameter space spanned by $\lambda_0, \lambda_1 $  in Fig. \ref{fig:OurHDifferences} (a) and (c), respectively. The difference is maximal in the MZM/MPM phase at points where the offset time variance,
\begin{equation}
V_\alpha = \Big\langle \big(\gamma_\alpha(t)-\langle \gamma_\alpha(t) \rangle \big)^2\Big\rangle,
\end{equation}
 of the function controlling the resonance width $\gamma_\alpha(t)$ is greatest, as captured in Fig. \ref{fig:OurHDifferences} (b) and (d).  	
Fig. \ref{fig:OurHDifferences} (e-h) show the  stroboscopic and DC conductance traces  taken at values of  $\lambda_0 $ and $\lambda_1 $ where the difference function $D_{0/\pi}$ is maximal. This difference, although small, is persistent in the limit $\Gamma \to 0$. Panel (j) shows the energy-integrated difference between the stroboscopic and DC conductance. Notably, this is quantitatively well captured by the figures of merits $D_0$ and $D_\pi$ obtained from the discrepancy between conductances at the resonances. Upon increasing the coupling to the external leads, the agreement between the conductance resonances in the two coupling configurations breaks down and the stroboscopic and DC conductance are increasingly different [cf. panel (i)].

\subsection{Sudden switching between Hamiltonians within the trivial topological phase}
\begin{figure}
	\centering
	\includegraphics[width=0.5\textwidth]{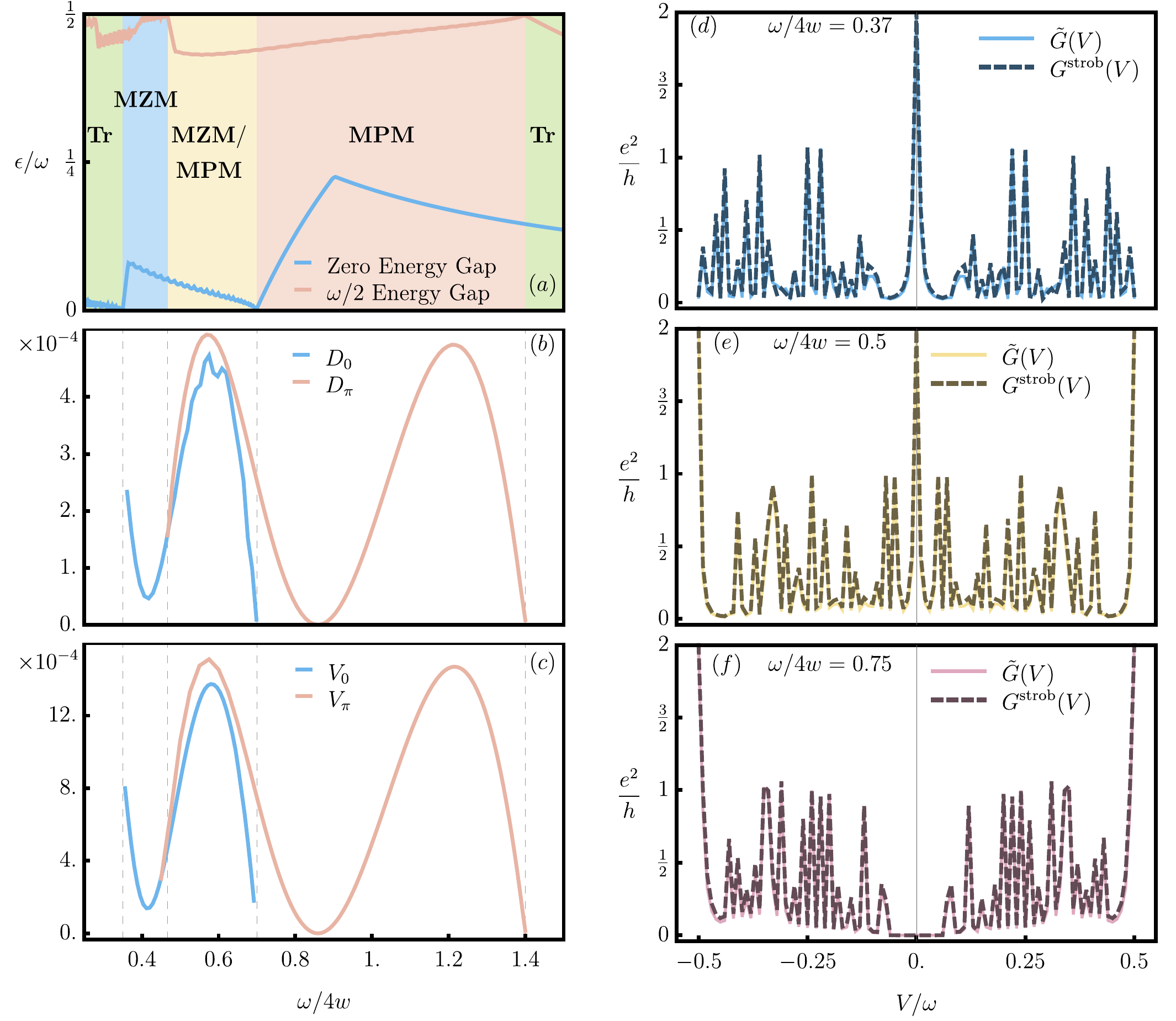}
	\caption{(a) Variation of the bulk quasi energy gaps around $\epsilon=0$ and $\epsilon=\pi$ over a range of driving frequencies $\Omega$. The possible phases over this range are denoted: Tr=Trivial, MZM=Zero-modes only, MPM=$\pi$-modes only and MZM/MPM=Both Zero and $\pi$ modes. (b) Illustrates the behaviour of the difference functions for the zero ($D_0$) and $\pi$ ($D_\pi$) mode resonances over this range of frequencies and (c) the corresponding behaviour of the time variance of $\gamma_i(t)$ dictating this difference. (d-f) Conductance profiles for selected driving frequencies comparing the measured conductance summed over sidebands with the stroboscopic construction. All data was obtained using a chain length of $n=70$ and coupling strength $\Gamma_{L/R}/\omega = 0.04$ }
	\label{fig:PRL}
\end{figure}

As a second example we consider a different driving protocol that consists of instantaneously switching between two topologically trivial Hamiltonians, which differ by the value of the chemical potential $\mu(t)$ \cite{Kundu2013}. 
\begin{eqnarray}
	\label{PRL Kitaev Ham 1}
	H_0&=&H(w, \Delta,\mu_0)\\
	\label{PRL Kitaev Ham 2}
	H_1 &=&H(w, \Delta,\mu_1)
\end{eqnarray}
such that $\mu_{0,1} > w/2 $.  Here the phase diagram is spanned by changing the driving frequency  $\Omega/w$, see Fig. \ref{fig:PRL} (a).

In Fig. \ref{fig:PRL} (d-f) we compare the conductance traces  for selected driving frequencies summed over side bands with the stroboscopic conductance. As in the previous case, we find that the conductance calculated using the sum rule shows  good agreement with  the stroboscopic fictitious conductance. In panels (b) and (c) we see that, once again, the time variance of the function $\gamma_i(t)$ captures the behaviour of both differences between the conductances at the zero and $\pi$ Majorana peaks as a function of the driving frequency.  

While the difference between the stroboscopic and DC conductance appears to be small across the whole voltage range and is well captured by the difference between the width of the resonant peaks, some aspects of the differences between the two conductances are protocol-dependent. The maximum difference in the second protocol is roughly two orders of magnitude smaller than in areas of maximum difference of the first protocol. This is due to the time dependence of the function $\gamma(t)$, which is  dictated by the how the structure of the eigenstates of the Floquet operator depend upon the offset time between the stroboscopic coupling and the start of the driving cycle. In particular it depends upon the contribution to the eigenstates on the fermionic sites at the ends of the chain, which are coupled to the external leads. At the points of maximum difference in the first protocol, the eigenstates are highly localized on the final site when $t=0$ but as the offset time increases this localization is shifted almost entirely to the second site and then back to the first at $t=0.5T$. This leads to significant time dependence of $\gamma(t)$ and consequently a difference between the two conductance quantities. In the second protocol, the eigenstates are more evenly distributed throughout the entire chain and the contribution on the end sites seems to depend little on the variation of the offset time $t$.

\section{Conclusions}
In conclusion, we have established a connection between the experimentally accessible Floquet scattering matrix for  driven topological systems and  a related scattering matrix for the case of pulsed coupling to external leads  \cite{Fulga2016}. The comparison between the two provides a platform from which to explore signatures of Floquet topological phases in transport properties such as differential conductance, building upon the relationships between scattering matrix invariants and topological phase classification established for non-driven systems.

We have compared the transport properties of the two coupling configurations by analyzing the DC conductance summed over Floquet side bands and the period-averaged stroboscopic conductance of pulsed coupling to the leads, both showing conductance peaks quantized at height of $2e^2/h$.  We have shown that in general the difference between the two is captured by the different widths of the resonant peaks in the limit of small coupling to the external leads.

We have demonstrated  this relation in a specific example of a periodically driven Kitaev chain,  considering  two different driving protocols:  A driving protocol which  instantaneously switches between Hamiltonians at the sweet spots of the trivial and topological phases, and a driving protocol which instantaneously switches between two Hamiltonians within the same topologically trivial phases.
Both protocols show the emergence of four distinct topological phases which can be accessed via  tuning of the Hamiltonian parameters. Each of the three non trivial  phases is characterized by the existence of Majorana zero or $\pi$ modes. These topological phases can be characterized by scattering matrix invariants formulated in terms of  a gedanken pulsed scattering experiment \cite{Fulga2016}. The three non trivial phases are also characterized by DC conductance signatures which, when summed over Floquet side bands, result in  conductance peaks quantized at height of $2e^2/h$, at values of external bias corresponding to the energy of the Majorana modes characteristic of the given phase.
We have shown  that the difference between the DC experimentally accessible conductance  and the  conductance obtained from the gedanken pulsed measurement is reflected in the width of the zero and $\pi $ conductance peaks. We have studied the dependence  of the  difference function on the physical parameter space in the two protocols and found that  generically the discrepancy  is larger when the  zero mode weight at the end of  the chain depends strongly on  the offset time between the driving cycle and the pulsed coupling period.  

 
 Although we have focused here upon the example of a driven Kitaev chain, our methodology could be applied to any open, periodic system to explore the connection between experimentally accessible transport features and Floquet topological phases.     
\section{Acknowledgements} We acknowledge fruitful discussion with 
Babak Seradjeh. D.M. acknowledges  support from the Israel Science foundation (grant No. 8114881).

\bibliography{References.bib}
	
	\newpage
	
	\appendix
	
	\onecolumngrid

	\section{Computation of the current in terms of $S_t(E)$}
	\label{app:appendice}
	
	We derive here Eqs. \ref{eq:corrente}. Our starting point is Eq. \ref{eq:current1},
	
	We use the time periodicity in the Fourier expansion of the scattering states operators:
	\begin{eqnarray}
	a(lT+t) = \int dE a^F(E) e^{-i E(lT+t)},
	\end{eqnarray}
	where $a^F(E) = \sum_l  a(lT+t) e^{i E(lT+t)}$ and $\langle {a_\alpha^F}^\dag (E') a_\alpha^F (E )\rangle = f_\alpha^F (E) \delta(E-E')$
	to get
	\begin{eqnarray}
	\nonumber
	I_\alpha(lT+t)=\int dE &&\left\{ \sum_{\beta} \sum_{m',m''} S_{\alpha,\beta}^*(t+lT,t+m'T)S_{\alpha,\beta}(t+lT,t+m''T)f_\beta^F (E)  e^{i E (m'T-m''T) }-f_\alpha^F (E)\right\}.
	\end{eqnarray}
	This expression is simplified  by plugging in the Fourier expansion of the scattering matrix
	and	using $e^{i n \Omega T }=1 $ and $\sum_{m'} e^{i(ET)m'} = \frac{2 \pi}{T}\delta(E)$ to arrive at
	\begin{eqnarray}
	I_\alpha(lT+t)&=&\int dE \left\{ \sum_{\beta} \sum_{n',n''} S_{\alpha,\beta}^*(E,E_{n'}) S_{\alpha,\beta}(E,E_{n''})f_\beta^F (E)e^{i (n''-n')\Omega  t}-f_\alpha^F (E) \right\}.
	\end{eqnarray}
	
	Finally, the current is expressed in terms of a  stroboscopic scattering  matrix 
	\begin{eqnarray}
	\nonumber
	I_\alpha(lT+t)=\!\!\int \!\!dE\sum_{\beta} \left[ |S^{F}_{\alpha,\beta,(t)}(E)|^2f_\beta^F(E)\right]-f_\alpha^F (E), \label{eq:corrente-app}\\
	\end{eqnarray}
	where
	\begin{eqnarray}
	\sum_n S_{\alpha,\beta}(E,E_{n})e^{i n\Omega  t} = S^{F}_t(E).
	\end{eqnarray}
	
	Once we have identified the relation between  scattering matrix of the  full time dependent problem  the stroboscopic  scattering matrix of  the fictitious problem,  we can express the  actual current in terms of $S_t (E)$. For this we manipulate Eq. \ref{eq:corrente-app} along the lines of Ref. \cite{Moskalets2002b}:
	\begin{eqnarray}
	\nonumber
	I_\alpha  &=& \frac{e}{h} \int_0^\infty \sum_{\beta\neq \alpha} \sum_{n}\left\{  |S_{\alpha \beta}(E_n,E)|^2f_\beta(E)\right.\\
	&& \left.-|S_{\beta\alpha}(E_n,E)|^2f_\alpha(E)\right\}.
	\end{eqnarray}
	We also note that 
	\begin{eqnarray}\label{MosvsFulga}
	\nonumber
	\sum_n S_{\alpha,\beta}(E,E_{n})e^{i n\Omega  t}  &=&\sum_n S_{\alpha,\beta}(E_{-n},E)e^{i n\Omega  t} \\
	&=& \sum_m S_{\alpha,\beta}(E_{m},E)e^{-i m\Omega  t}  = S^{F}_t(E)
	\end{eqnarray}
	where in the first equality we have redefined our energy to be $E\rightarrow E+n\Omega $ and in the second we have changed the sum to be over $m = -n $. We can also use the fact that eq. \eqref{MosvsFulga} defines a discrete FT s.t.:
	\begin{eqnarray}\label{MosvsFulga}
	S_{\alpha,\beta}(E_n,E)=\frac{2\pi}{T} \int_0^T S^{F}_t(E)e^{in\Omega t}.
	\end{eqnarray}
	Using this expression we can write: 
	\begin{eqnarray}
	\nonumber 
	I_\alpha  = \frac{e}{h} \int_0^\infty dE \left(\frac{2\pi}{T}\right)^2\int_0^T dt dt' \sum_{\beta\neq \alpha} \sum_{n}\left\{  S^{F*}_{\alpha \beta,(t)}(E)S^{F}_{\alpha \beta,(t')}(E)f_\beta(E) -S^{F*}_{\beta\alpha,(t)}(E)S^{F}_{\beta\alpha,(t')}(E)f_\alpha(E)\right\}e^{in\Omega (t'-t) }.\\
	\end{eqnarray}
	We perform the sum over $n$ which results in the final expression (cf. Eq. \ref{eq:corrente})
	\begin{eqnarray}
	\nonumber 
	I_\alpha & =& \frac{e}{h} \int_0^\infty dE \sum_{\beta\neq \alpha}   \left(\int_0^T \frac{2\pi dt}{T}|S^{F}_{\alpha \beta,(t)}(E)|^2\right)f_\beta(E) \\
	&-& \left(\int_0^T \frac{2\pi dt}{T}|S^{F}_{\beta\alpha,(t)}(E)|^2\right)f_\alpha(E).
	\end{eqnarray}

\end{document}